\documentclass[preprint,12pt]{elsarticle}

\usepackage{bm}
\usepackage{booktabs}
\usepackage{graphicx} 
\RequirePackage{amsmath}
\RequirePackage{amsthm}
\RequirePackage{dsfont}
\RequirePackage{amssymb}
\usepackage{float}
\usepackage{subcaption}
\usepackage{colortbl,color,soul}
\usepackage{algorithm}
\usepackage{eurosym}
\usepackage{algpseudocode}
\RequirePackage[usenames,dvipsnames,table]{xcolor}
\usepackage[%
  colorlinks = true,
  citecolor  = Black,
  linkcolor  = Black,
  urlcolor   = RoyalBlue,
  unicode,
  ]{hyperref}
\usepackage{cleveref}
\usepackage{supertabular}
\usepackage[intoc]{nomencl}  
\usepackage{mathtools} 
\usepackage{indentfirst}
\usepackage{nicefrac}
\allowdisplaybreaks

\input{Acronyms}
\usepackage[intoc]{nomencl}
\usepackage{xstring}
\usepackage{xpatch}
\usepackage{eurosym}

\patchcmd{\thenomenclature}
  {\leftmargin\labelwidth}
  {\leftmargin\labelwidth\itemindent 1em }
  {}{}
\newcommand{\nomenclheader}[1]{%
  \item[\hspace*{-\itemindent}\normalfont\bfseries#1]}
\renewcommand\nomgroup[1]{%
  \IfStrEqCase{#1}{%
   {B}{\nomenclheader{Bernstein Polynomials:}}
   {C}{\nomenclheader{CUC:}}
   {E}{\nomenclheader{MILP CUC:}}
   {F}{\nomenclheader{CFCUC:}}
  }%
}

\makenomenclature

\nomenclature[B]{\(f(\tau)\)}{a function of $\tau$}
\nomenclature[B]{\(\tau\)}{time variable}
\nomenclature[B]{\({B_n}^f(\tau)\)}{bernstein polynomial of order $n$ for function $f$}
\nomenclature[B]{\({C_b}^f\)}{$b$th Bernstein coeffiecient to represent $f$}
\nomenclature[B]{\(n\)}{degree of Bernstein polynomial}
\nomenclature[B]{\(b\)}{index of the Bernstein coefficient [$\in\{0,\dots,n-1\}$]}
\nomenclature[B]{\(\beta_{b,n}(\tau)\)}{Bernstein basis function}
\nomenclature[B]{\({B^{'}_n}^f(\tau)\)}{differential of Bernstein polynomial}

\nomenclature[C]{\(\text{suc}(.)\)}{start-up costs [k\euro]}
\nomenclature[C]{\(\text{gc}(.)\)}{generation costs [k\euro]}
\nomenclature[C]{\(u_i\)}{commitment variable [$\in\{0,1\}$}
\nomenclature[C]{\(v_i\)}{start-up variable [$\in\{0,1\}$}
\nomenclature[C]{\(w_i\)}{shut-down variable [$\in\{0,1\}$}
\nomenclature[C]{\(p\)}{power generation variable [MW]}
\nomenclature[C]{\(p'\)}{differential of power [MW/s]}
\nomenclature[C]{\(t\)}{index of time intervals}
\nomenclature[C]{\(T\)}{set of all time intervals}
\nomenclature[C]{\(i\)}{index of generators}
\nomenclature[C]{\(\mathcal{I}\)}{set of all generators}
\nomenclature[C]{\(s\)}{alias index for time intervals}
\nomenclature[C]{\(\text{UT}\)}{minimum up-time of generators [hours]}
\nomenclature[C]{\(\text{DT}\)}{minimum down-time of generators [hours]}
\nomenclature[C]{\(\overline{\mathcal{P}}_i\)}{maximum power output of generator $i$ [MW]}
\nomenclature[C]{\(\underline{\mathcal{P}}_i\)}{minimum power output of generator $i$ [MW]}
\nomenclature[C]{\(\overline{\mathcal{R}}_i\)}{maximum ramp-up of generator $i$ [MW/h]}
\nomenclature[C]{\(\underline{\mathcal{R}}_i\)}{maximum ramp-down of generator $i$ [MW/h]}
\nomenclature[C]{\(r\)}{online reserve power variable [MW]}
\nomenclature[C]{\(p^{\text{\tiny RES}}\)}{RES generation variable [MW]}
\nomenclature[C]{\(\overline{\mathcal{D}}\)}{Demand [MW]}

\nomenclature[E]{\(\bm{D}\)}{vector of load coefficients}
\nomenclature[E]{\(\bm{\beta}\)}{vector of Bernstein basis functions}
\nomenclature[E]{\(\bm{P}\)}{vector of power coefficients}
\nomenclature[E]{\(\bm{P}^{\text{\tiny RES}}\)}{vector of RES coefficients}
\nomenclature[E]{\(\bm{P}^{\text{curt}}\)}{vector of curtailment coefficients}
\nomenclature[E]{\({C_{b,t,i}}\)}{$b$th Benstein coefficient at time $t$ of unit $i$}
\nomenclature[E]{\(\text{SU}^T_i\)}{tart-up time duration of unit $i$ [h]}
\nomenclature[E]{\(\text{SD}^T_i\)}{shut-down time duration of unit $i$ [h]}
\nomenclature[F]{\(\ell\)}{index of the lost unit}
\nomenclature[F]{\(\Delta f_{\text{crit}}^{'}\)}{critical RoCoF [Hz/s]}
\nomenclature[F]{\(f_0\)}{nominal frequency [Hz]}
\nomenclature[F]{\(H_\ell\)}{available weighted inertia after outage of $\ell$ [MW.s]}
\nomenclature[F]{\(\bm{R}\)}{vector of reserve coefficients}
\nomenclature[F]{\(D\)}{load damping factor [\%/Hz]}
\nomenclature[F]{\(\Delta f^{\text{ss}}_{\text{crit}}\)}{critical steady-state frequency [Hz]}
\nomenclature[F]{\(f^{\text{nadir}}\)}{frequncy nadir [Hz]}
\nomenclature[F]{\(T^g\)}{delivery time of the units [s]}
\nomenclature[F]{\(\alpha\)}{coefficients in the linear model}

\journal{IJEPES}

\begin{document}

\begin{frontmatter}



\title{Data-Driven Continuous-Time Framework for Frequency-Constrained Unit Commitment}


\author[inst1]{Mohammad Rajabdorri}
\author[inst1]{Enrique Lobato}
\author[inst1]{Lukas Sigrist}

\affiliation[inst1]{organization={Instituto de Investigación Tecnológica (IIT), Escuela Técnica Superior de Ingeniería (ICAI), Universidad Pontificia Comillas},
            city={Madrid},
            country={Spain}}

\author[inst2]{Jamshid Aghaei}
\affiliation[inst2]{organization={School of Engineering and Technology, Central Queensland University},
            city={Queensland},
            country={Australia}}

\begin{abstract}
The conventional approach to solving the unit commitment problem involves discrete intervals at an hourly scale, particularly when integrating frequency dynamics to formulate a frequency-constrained unit commitment. To overcome this limitation, a novel continuous-time frequency-constrained unit commitment framework is proposed in this paper. In this approach, Bernstein polynomials represent continuous variables in the unit commitment problem and enable the calculation of frequency response-related metrics such as the rate of change of frequency, quasi-steady-state frequency, and frequency nadir, and the corresponding continuous-time constraints are introduced.
Notably, startup and shut-down trajectories are meticulously considered, transforming the formulation into a fully continuous-time model and simplifying constraints related to variable continuity.
To address the complexities associated with integrating the obtained non-linear frequency nadir constraint into a mixed-integer linear problem, an alternative data-driven frequency nadir constraint is proposed, which accurately constrains frequency nadir deviations throughout the time interval. To validate the proposed model, it is applied to the real-life network of the Spanish Island of La Palma. The results demonstrate the effectiveness of the proposed formulation, indicating that the model is solved timely while mitigating the impact of intra-hour real-time power fluctuations on system frequency.
\end{abstract}




\end{frontmatter}


\renewcommand*{\glsclearpage}{\clearpage}
\printglossary[type=\acronymtype,style=mylong]
\printnomenclature

\section{Introduction}

Thanks to the recent progress, the short-term load and \gls{RES} forecasts are more accurate than ever \cite{lin2021spatial,jiao2021adaptive,duan2022novel,li2022short}. Although forecasting methods are improving rapidly, the forecasted quantities should be discretized into energy blocks (usually 1-hour) when used for operation planning. The use of discrete forecasted load and \gls{RES} forecasts in \gls{UC} yields discrete generation and reserve schedules. The use of such discrete intervals is a source of error by neglecting sub-hourly changes in demand and \gls{RES} for instance. As it is not physically feasible to instantaneously ramp up/down at hourly intervals, \glspl{ISO} would employ different ways to make that possible. For example, California \gls{ISO} instructs a 20-min linear ramp across the hourly intervals \cite{caiso2018business}. The deviation of the forecasted quantities from the actual quantities will be treated in subsequent planning and real-time operation steps, leading to additional costs. These variations can be significant in island power systems, where \gls{RES} generation is geographically located in a few areas, for instance, and can cause variations in the sub-hourly generation dispatch with respect to the expected one. 

Low-inertia power systems in general and island power systems in particular are sensitive to active power disturbances, causing significant frequency variations. \cite{zhang2023security} highlights the importance of inertia in power grids, especially with the increase in \gls{RES} and DC transmission. The most challenging disturbances are generation outages in islands. \gls{UC} formulations considering post-disturbance frequency dynamics have been proposed in the literature \cite{badesa2019simultaneous,trovato2018unit,zhou2023frequency,lagos2021data,zhang2020approximating,sang2023conservative,rajabdorri2022robust,rajabdorri2023inclusion}. Ensuring the adequacy of frequency response is more crucial than ever, as the system inertia is declining because of the growing implementation of non-synchronous \gls{RES}. If the frequency response is not sufficient when a grid-related event happens, load shedding happens, and if not done rapidly, cascading events leading to a blackout will occur. This emphasizes the
importance of identifying and maintaining frequency deviations \cite{stankovski2023power}.

In \cite{badesa2019simultaneous} a stochastic \gls{UC} method optimizes various frequency services concurrently, specifically tailored for low inertia systems featuring substantial \gls{RES} integration. Employing scenario trees generated through a quantile-based scenario generation approach, the stochastic model addresses the challenge of linearizing frequency nadir. The technique involves an inner approximation method for one side of the equation and utilizes binary expansion, linearized through the big-M technique, for the other side. \cite{trovato2018unit} present a linear formulation in their study, addressing the \gls{UC} problem by incorporating details about inertial response and system frequency response. This formulation ensures adequate ancillary service provision to prevent under-frequency load shedding after the biggest outage. To linearize the frequency nadir constraint, they calculate first-order partial derivatives of its equation for higher-order non-linear variables, ultimately representing the frequency nadir through a set of piecewise linearized constraints. In \cite{zhou2023frequency} the differential-algebraic equation of the frequency response is piecewise approximated using Bernstein polynomials and then linearized, to reflect frequency nadir. The \gls{FCUC} that is proposed in \cite{jiang2022coordinative} integrates an emergency frequency control strategy utilizing the transient overload capabilities of \gls{HVDC} systems for inertial response. Additionally, it introduces an AC-DC primary response model considering \gls{HVDC} decline rates, facilitating optimized AC-DC secondary response solutions. \cite{han2024robust} proposes a method for obtaining frequency constraints in low-inertia grids, focusing on the optimal allocation of distributed energy storage systems for locational frequency security. In \cite{li2023novel} a dynamic frequency response model is analyzed to derive the relation between frequency nadir and reserve power. In the day-ahead scheduling stage, optimal power reserve for distributed generators is determined to satisfy the primary and secondary frequency regulation.

In contrast to deriving analytical formulas from the swing equation employed in \cite{badesa2019simultaneous,trovato2018unit,jiang2022coordinative}, \cite{lagos2021data} adopts a data-driven multivariate optimal classification trees technique. They propose a robust formulation to address load and \gls{RES} uncertainties. The classification tree is independently solved as a \gls{MILP} problem. \cite{zhang2020approximating} contribute a dynamic model for generating training data, subsequently utilized in training a deep neural network. The trained neural networks are structured for integration into an \gls{MILP} problem, facilitating the development of a frequency nadir predictor for application in the \gls{UC} problem. In \cite{sang2023conservative} a conservative sparse neural network is used to learn the frequency dynamics. Then the learned model is encoded as \gls{MILP} to be incorporated into the \gls{UC} problem. A robust formulation with data-driven frequency constraint is proposed in \cite{rajabdorri2022robust}, and in \cite{rajabdorri2023inclusion} frequency nadir constraint is obtained, using a synthetically generated dataset.

An unwritten assumption in these studies is that power is delivered in hourly constant blocks and the outage of the unit during an hour yields the same frequency response. 
In reality, outages can happen at any moment throughout the time interval, and power outputs are subjected to intra-hour changes. When an outage happens depending on the instantaneous power output of the units, the frequency response might be different. To be able to restrain the frequency response continuously throughout the time interval, we should formulate the \gls{UC} problem as a continuous-time model. In \cite{parvania2015unit} a continuous-time model based on Bernstein polynomials is introduced. As the output of \gls{CUC} presented in \cite{parvania2015unit} is a continuous curve, it also captures the ramping of the units. So this \gls{CUC} can ensure there are enough ramping resources to handle the evergrowing variabilities in the power system and larger ramping events. The adequacy of ramping resources should be ensured, particularly due to the addition of solar power, which exacerbates the need for fast ramping units in the early morning and early evening hours \cite{scaglione2016continuous}. 
Bernstein polynomials are further utilized in \cite{bagherinezhad2020continuous} to develop a novel continuous-time optimization model for co-optimizing balancing energy and flexible ramp products in real-time power system's operation. These polynomials are employed to model the continuous trajectories of up and down flexible ramp products, ensuring the deliverability of ramping capacity. 
And in \cite{zhou2023function} by employing Bernstein polynomials, the function-space optimization is converted into \gls{MILP} for tractable calculation. 
While employing Bernstein polynomials instead of the traditional discrete \gls{UC} formulation may introduce additional computational burden and potentially impact scalability, it is not a concern in this context. This paper's emphasis is on low-inertia island power systems, which tend to be relatively small in scale.

The \gls{CUC} formulation of \cite{parvania2015unit} is employed as a starting point for the proposed \gls{CFCUC}. Frequency response metrics like rate of change of frequency, quasi-steady-state frequency, and frequency nadir are obtained from the swing equation and defined as Bernstein polynomials in this paper. In the case of frequency nadir, the obtained term is highly non-linear, which makes it hard to calculate and apply to the \gls{MILP} problem in hand, even after considering the simplifying assumptions that are suggested later in the methodology section. To avoid that, a data-driven approach is utilized to acquire a linear frequency nadir constraint similar to \cite{rajabdorri2023inclusion}. Linear regression is employed to drive the frequency nadir constraint from this synthetic dataset in terms of Bernstein coefficients. Data-driven methods are proven to be effective in describing frequency dynamics in the \gls{FCUC} problem \cite{lagos2021data,zhang2020approximating,sang2023conservative,rajabdorri2022robust,rajabdorri2023inclusion}. It's revealed in the results section that the simplifying assumption and the learning procedure can continuously restrict the frequency nadir deviation very accurately throughout the time interval.

\begin{figure}[!htbp]
    \centering
    \includegraphics[width=0.9\linewidth]{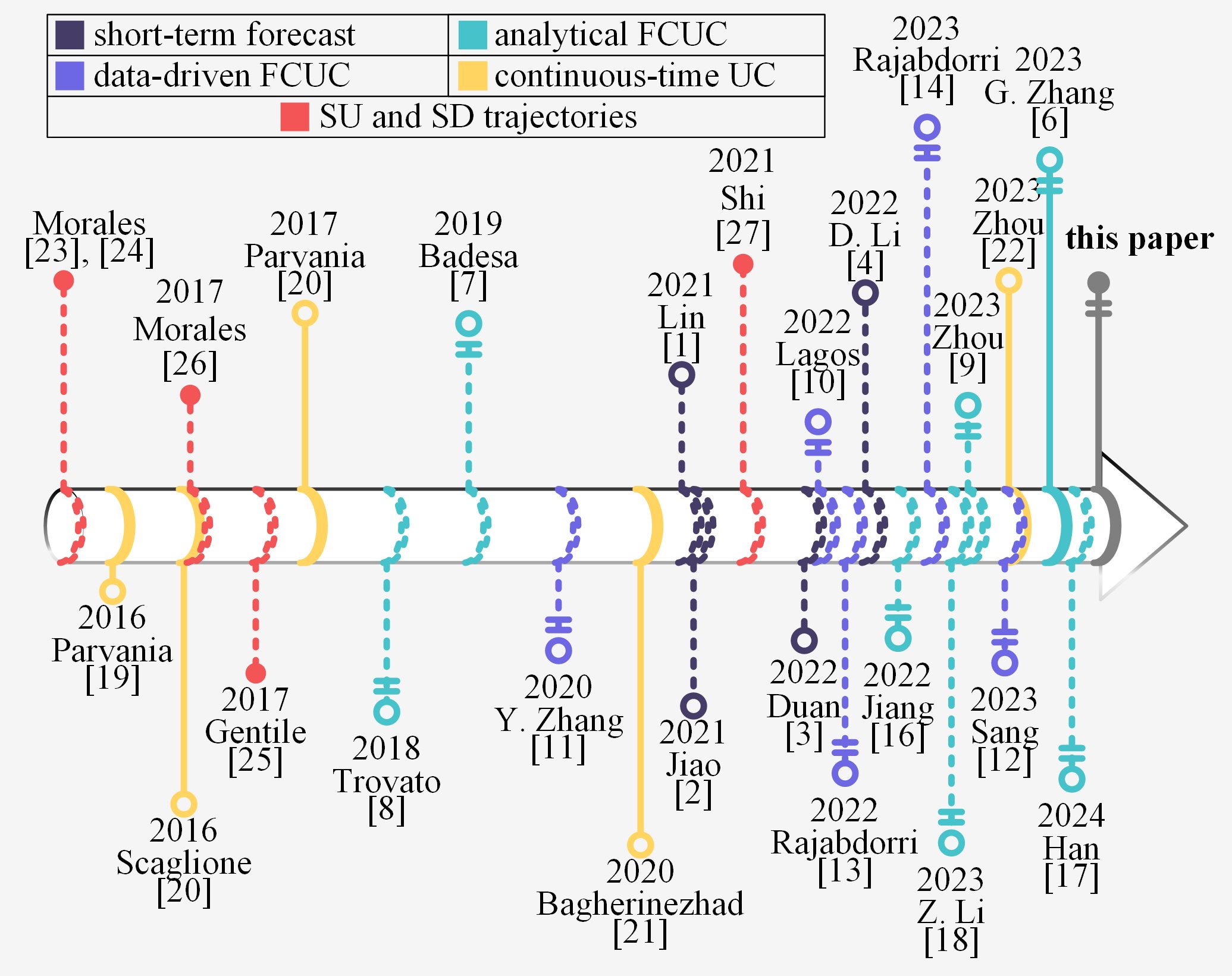}
    \caption{Summary of the cited studies.}
    \label{fig:timeline}
\end{figure}

Another useful addition to the proposed \gls{CFCUC} formulation is \gls{SU} and \gls{SD} trajectories, which are ignored in \cite{parvania2015unit}. Power trajectories during \gls{SU} and \gls{SD} are mostly neglected in the \gls{UC} formulation. However, units are providing energy and ramp during the \gls{SU} and \gls{SD} process. Optimally scheduling these trajectories is economically and technically favorable and is studied in \cite{morales2013mip,morales2012tight,gentile2017tight,morales2017hidden}. \cite{morales2013mip} claims that the traditional \gls{UC} may result in false estimations of the ramping capability of the units. Then, it presents a linear continuous schedule of generation that better fits the load profile and considers \gls{SU} and \gls{SD} trajectories.
\cite{shi2021optimal} analyzes the potential of short-time \gls{SU} and \gls{SD} regulation modes of thermal units to regulate system peak load. An optimal scheduling model is proposed, which includes an improved \gls{UC} model incorporating \gls{SU} and \gls{SD} trajectories of thermal units.
We particularly consider \gls{SU} and \gls{SD} trajectories because (1) the \gls{CFCUC} model will be fully continuous. (2) The zero-order continuity (which ensures the continuity of the scheduled power outputs) and the first-order continuity (which ensures the continuity of the derivative of the power outputs, i.e. the continuity of ramping) of Bernstein polynomials between intervals will be preserved, even when the start-up or shut-down of the units. (3) As claimed in \cite{morales2012tight} the model will be tighter and faster, which is very useful to alleviate the computational burden of the proposed \gls{CFCUC}. The cited studies are summarized in \cref{fig:timeline}. References with discrete formulation are shown dashed and the ones that use Bernstein polynomials with continuous lines. The ones that consider \gls{SU} and \gls{SD} trajectories with a solid bullet, and the ones that consider frequency constraints with =.

The contributions with respect to the previous studies as follows.
\begin{itemize}
    \item The frequency-related metrics are defined as continuous-in-time Bernstein polynomials, and the respective constraints are applied to the \gls{CUC} problem.
    \item The obtained frequency nadir equation based on Bernstein polynomials is highly non-linear. A data-driven approach is suggested to drive a linear frequency nadir constraint. The dataset is generated with a synthetic procedure, introduced in \cite{rajabdorri2023inclusion}. Simplifying assumptions are offered, to calculate the frequency nadir and label the sample data points.  
    \item \Gls{SU} and \gls{SD} trajectories are included in the \gls{CUC} formulation to improve its solution run-time and general representation.
\end{itemize}

The rest of the paper is organized as follows. The methodology is presented in \cref{sec:meth} which includes an introduction to the required basic of Bernstein polynomial (in \cref{sec:ber}), a general formulation of \gls{CUC} (in \cref{sec:cuc}), and the proposed \gls{FCUC} (in \cref{sec:CFCUC}). The obtained results are demonstrated in \cref{sec:results}. Finally, the conclusions are drawn in \cref{sec:conc}.

\section{Methodology}\label{sec:meth}

As mentioned before, the continuous variables in \gls{CUC} are defined as Bernstein polynomials. In this section, first, the necessary fundamentals about the Bernstein polynomials to define the \gls{CUC} are described. Then it's used to build the general \gls{CUC} and the proposed \gls{CFCUC} models.

\subsection{Bernstein Polynomials}\label{sec:ber}

\subsubsection{Function Approximation} 
\begin{figure}[!htbp]
    \centering
    \includegraphics[width=1\linewidth]{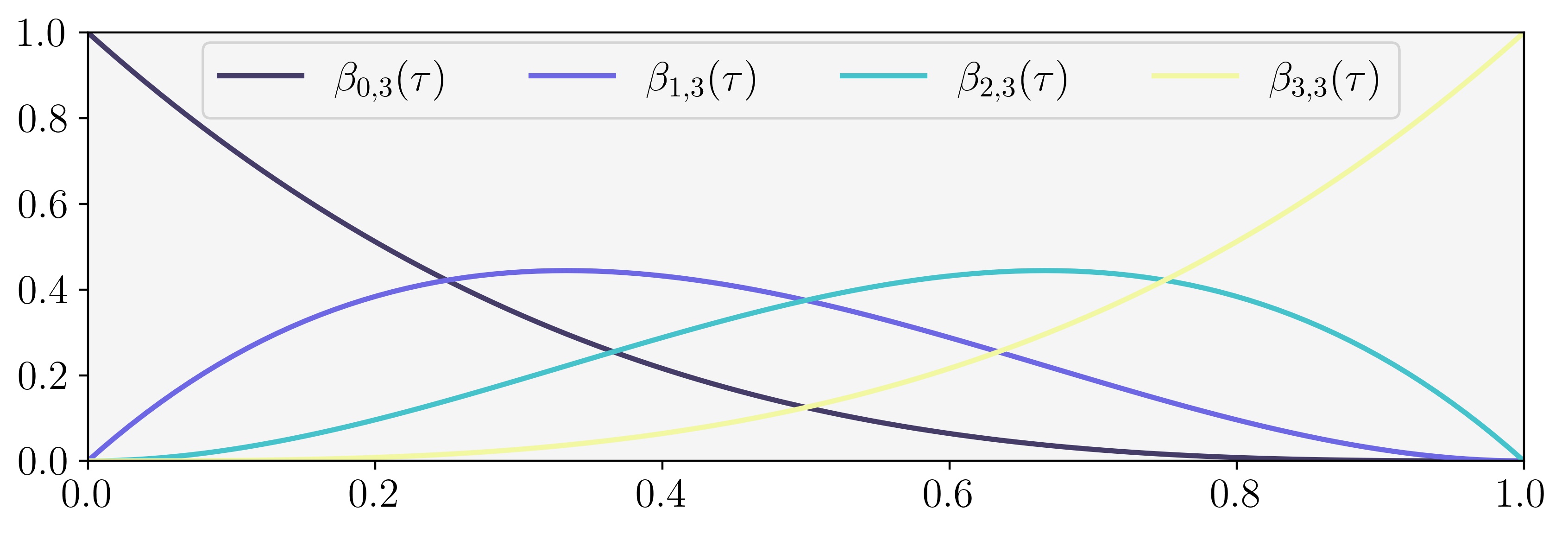}
    \caption{Bernstein basis functions of order 3.}
    \label{fig:ber_polys}
\end{figure}

Bernstein's function can approximate curves with a linear combination of polynomials. One advantage of using polynomials is that they can be calculated very quickly algebraically. For a function $f(\tau)$, defined on the closed interval ($\tau\in[0,1]$), the expression ${B_n}^f(\tau)$ is called the Bernstein polynomial of order $n$ for the function $f(\tau)$,
\begin{equation}\label{eq:ber_poly}
    {B_n}^f(\tau)=\sum_{b=0}^{n}{C_b}^f\times \beta_{b,n}(\tau),
\end{equation}
Where ${C_b}^f$ is the Bernstein coefficient and for $f(\tau)$ is equal to $f\big(\nicefrac{b}{n}\big)$. ${B_n}^f(\tau)$ is a polynomial in $\tau$ of order $n$ and $\beta_{b,n}(\tau)$ are the Bernstein basis functions.
\begin{equation}
    \beta_{b,n}(\tau)=\binom{n}{b}\tau^b(1-\tau)^{n-b}
\end{equation}
If $f(\tau)$ is continuous on $[0, 1]$, it is proven that the following equation is true in $[0,1]$,
\begin{equation}
    \lim_{n\to\infty}{B_n}^f(\tau)=f(\tau)
\end{equation}
So if the order on $n$ is sufficiently high ${B_n}^f(\tau)\approx f(\tau)$. In this paper, Bernstein polynomials of order 3 are used to describe continuous variables in each interval. \Cref{fig:ber_polys} shows all Bernstein basis functions of order 3. This figure may help visualize how Bernstein polynomial functions approximate curves. It's straightforward to extend the definition of a Bernstein polynomial in $[0,1]$ to $[t,t+1]$.

\subsubsection{Differential of a Polynomial} Differentiating (\ref{eq:ber_poly}) with respect to $\tau$ leads to:
\begin{equation}\label{eq:ber_def}
    {B^{'}_n}^f(\tau)=n\sum_{b=0}^{n-1}\big({C_{b+1}}^f-{C_b}^f\big)\binom{n-1}{b}\tau^b(1-\tau)^{n-1-b}
\end{equation}
\Cref{eq:ber_def} is a Bernstein polynomial of order $n-1$. Here, ${C_{b+1}}^f-{C_b}^f$ is the difference of the first order $\Delta f(\nicefrac{b}{n})$ of the function $f(\tau)$ at $\tau=\nicefrac{b}{n}$ \cite{lorentz2012bernstein}.

\subsubsection{Convex Hull Property}\label{sec:chp}
A general property of Bernstein polynomials, as concluded from \cite{dierckx1995curve}, is that it satisfies the convex hull property. Each Bernstein polynomial of order $n$ consists of $n$ different terms, each of them is the production of a coefficient ${C_b}^f$ and a Bernstein basis function.
Convex hull property indicates that ${B_n}^f(\tau)$  will never be outside the convex hull of the control polygon formed by Bernstein coefficients for each $b$. That is, ${B_n}^f(\tau)$ is always between maximum and minimum coefficients.
\begin{equation}\label{eq:limit}
    \max_{\forall b}\big({C_b}^f\big)\leq {B_n}^f(\tau) \leq \min_{\forall b}\big({C_b}^f)\big)
\end{equation}
This property significantly helps, later when maximum and minimum generations and ramping constraints are derived. Because of this property and other features mentioned in \cite{parvania2016generation}, Bernstein polynomials are convenient to use for solving \gls{CFCUC} problem, and especially to define and implement the frequency nadir constraint in the proposed \gls{CFCUC}.

\subsubsection{Sum and Difference polynomials} \label{sec:sum_diff} The sum and difference of two polynomials with the same order can be performed by simply adding and subtracting their Bernstein coefficients, respectively \cite{kielas2022bernstein}.

\subsubsection{Order Elevation} \label{sec:degree_elev} If the order of polynomials does not match, the one with the lower order should be elevated until they have the same order. Elevating ${B_n}^f(\tau)$ means finding a Bernstein polynomial with higher order $m\;(m>n)$ such that ${B_m}^f(\tau)={B_n}^f(\tau)$. The new $b$th Bernstein coefficients after elevating ${B_n}^f(\tau)$ by one are calculated as,
\begin{equation}
    \frac{b}{n+1}{{C_{b-1}}^f}+(1-\frac{b}{n+1}){{C_{b}}^f}
\end{equation}
This can be done recursively to increase the order.

\subsubsection{Multiplying Polynomials}\label{sec:multi_poli} In case we need to multiply two Bernstein Polynomials of ${B_n}^f(\tau)$ and ${B_m}^g(\tau)$ that are defined on the same interval, the product will be of order $n+m$ and the $k$th coefficients of the product are given by
\begin{equation}
    {C_k}^{fg} = \sum_{j=\max(0,k-m)}^{\min(n,k)}\frac{{\binom{n}{j}}{\binom{m}{k-j}}}{{\binom{n+m}{k}}}{C_j}^f{C_{k-j}}^g
\end{equation}
where $k\in \{0,\dots,n+m\}$.

\subsection{CUC Formulation}\label{sec:cuc}

Solving the \gls{UC} problem determines the schedule of all generating units in the system by considering their constraints and supplying the forecasted load at minimum cost. A general \gls{CUC} formulation is expressed below \cite{parvania2015unit},
\begin{subequations}
\begin{align}
&\min_{u,p}\ \text{suc}\big(u(\tau)\big)+\text{gc}\big(p_{i}(\tau)\big) \label{eq:obj}
\\
&u_{i}(\tau)-u_{i}(\tau-1)=v_{i}(\tau)-w_{i}(\tau) &&\text{\scriptsize $\tau\in T,\;i\in\mathcal{I}$}\label{eq:bi1}
\\
&v_{i}(\tau)+w_{i}(\tau)\leq1 &&\text{\scriptsize $\tau\in T,\;i\in\mathcal{I}$}\label{eq:bi2}
\\
&\int_{\tau-\text{UT}_i+1}^{\tau}v_{i}(\tau)d\tau\leq u_{i}(\tau) &&\text{\scriptsize $\tau-\text{UT}_i+1\geq 0,\;i\in\mathcal{I}$}\label{eq:up_time}
\\
&\int_{\tau-\text{DT}_i+1}^{\tau}w_{i}(\tau)d\tau\leq 1-u_{i}(\tau)&& \text{\scriptsize $\tau-\text{DT}_i+1\geq 0,\;i\in\mathcal{I}$}\label{eq:down_time}
\\
&p_{i}(\tau)\geq \underline{\mathcal{P}}_i u_{i}(\tau) &&\text{\scriptsize $\tau\in T,\;i\in\mathcal{I}$}\label{eq:min_gen}
\\
&p_{i}(\tau)+r_{i}(\tau)\leq \overline{\mathcal{P}}_i u_{i}(\tau) &&\text{\scriptsize $\tau\in T,\;i\in\mathcal{I}$}\label{eq:max_gen}
\\
&p^{'}_{i}(\tau)\geq \underline{\mathcal{R}}_i &&\text{\scriptsize $\tau\in T,\;i\in\mathcal{I}$}\label{eq:down_ramp}
\\
&p^{'}_{i}(\tau)\leq \overline{\mathcal{R}}_i &&\text{\scriptsize $\tau\in T,\;i\in\mathcal{I}$}\label{eq:up_ramp}
\\
&\sum_{i\in\mathcal{I}}\big(p_{i}(\tau)\big)+p^{\text{\tiny RES}}(\tau)=\mathcal{D}(\tau) &&\text{\scriptsize $\tau\in T$}\label{eq:power_balance}
\end{align}
\end{subequations}
The objective is to minimize \cref{eq:obj}, where $\text{suc}(\cdot)$ is the start-up cost and $\text{gc}(\cdot)$ is the generation cost function. \Cref{eq:obj} is subjected to \cref{eq:bi1,eq:bi2,eq:up_time,eq:down_time,eq:min_gen,eq:max_gen,eq:down_ramp,eq:up_ramp,eq:power_balance}. \Cref{eq:bi1,eq:bi2} represent the binary logic of the \gls{UC} problem. \Cref{eq:up_time,eq:down_time} are the continuous representation of the units' minimum up-time and minimum downtime constraints. \Cref{eq:min_gen,eq:max_gen} are the minimum and the maximum power generation constraints respectively. \Cref{eq:down_ramp,eq:up_ramp} are ramp-down and ramp-up constraints. \Cref{eq:power_balance} is the power balance equation. This formulation leads to a complex and computationally intractable time-variant problem, that is, it is not straightforward to solve.

\subsection{MILP formulation of CUC}\label{sec:milp_cuc}

The following approximations and modifications are provided to convert this intractable problem into a \gls{MILP}. Doing so enables us to use \gls{MILP} techniques that are fairly fast and already developed.
The Bernstein approximation is used to fit the load curve and forecasted wind generation. In this method, all continuous curves are divided into one-hour intervals ($[t,t+1]$) and each interval is approximated by a Bernstein polynomial of order three. Higher-order polynomials can be used if more accuracy is needed. The output will be the optimized continuous-time generation of units, also represented by Bernstein polynomials of order 3. Thus the optimization variables are the Bernstein coefficients of the power output of the units. Ramping schedule of units is defined as the differentiation of generation. According to \cref{eq:ber_def}, if the generation is of order 3, ramping will be presented by Bernstein polynomials of order 2. It's assumed that the binary variables can only change value at the beginning of an hour, so they are not defined as Bernstein polynomials. That follows the traditional way of committing the units.

Different parts of \cref{eq:obj,eq:bi1,eq:bi2,eq:up_time,eq:down_time,eq:min_gen,eq:max_gen,eq:down_ramp,eq:up_ramp,eq:power_balance} are reformulated as follows to achieve a formulation that is representable as \gls{MILP}. In this formulation, the \gls{SU} and \gls{SD} power trajectories are also considered. The simplifying assumptions in this paper regarding the \gls{SU} and \gls{SD} trajectories are: 1) the duration of the \gls{SU} process and the \gls{SD} process is assumed to be $\text{SU}^T_i$ and $\text{SD}^T_i$, respectively and 2) the unit is at its minimum output power at the end (beginning) of the \gls{SU} (\gls{SD}) process. Note that the other assumption in \cite{morales2013mip, morales2012tight, gentile2017tight, morales2017hidden} regarding a linear variation of the power in time during \gls{SU} and \gls{SD} process is not needed here, as the Bernstein polynomials will describe the ramping.

\subsubsection{Load Approximation} 
\begin{figure}[!htbp]
    \centering
    \includegraphics[width=1\linewidth]{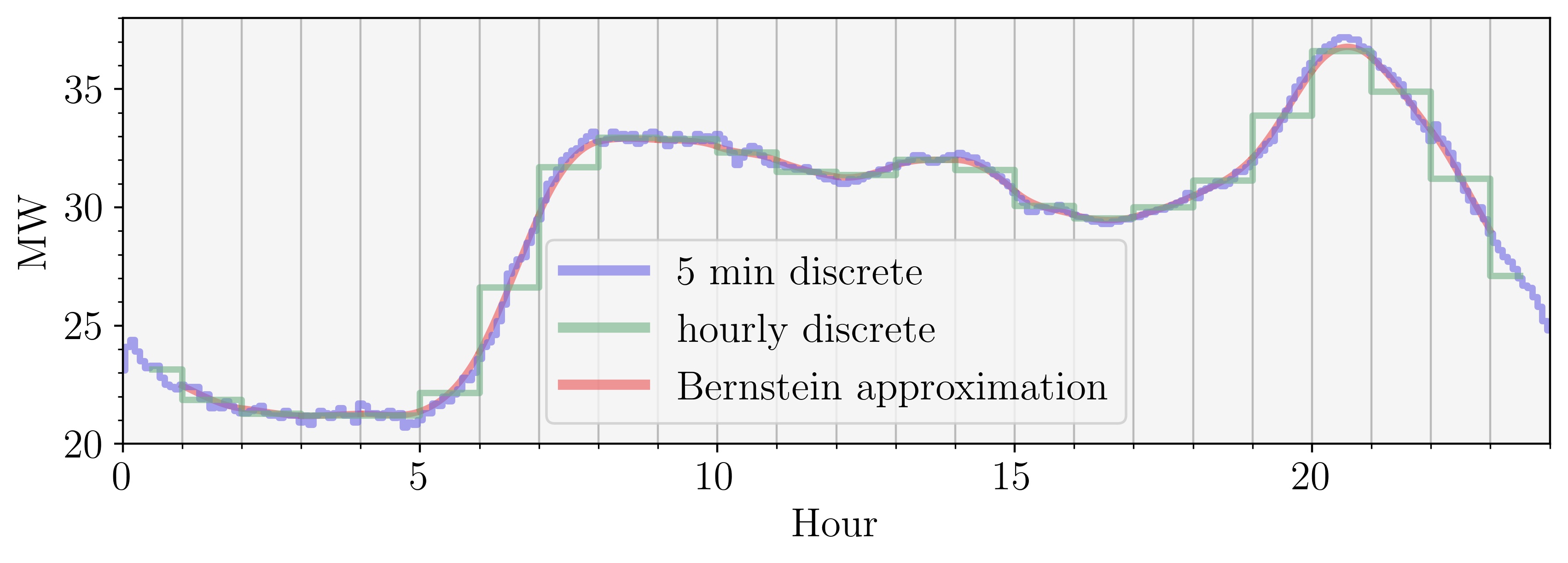}
    \caption{Hourly discrete, 5 min discrete, and Bernstein approximation of load.}
    \label{fig:load}
\end{figure}

Each hourly interval $[t,t+1]$ of the input load is approximated by a Bernstein polynomial of order 3. This approximation is more accurate than the hourly discrete approximation used in practice (see \cref{fig:load}, where the Bernstein polynomial almost perfectly follows the 5-min forecast).
Each interval is defined as the product of Bernstein coefficients and Bernstein basis functions as follows,
\begin{equation}
    \mathcal{D}(\tau)\approx {B_{3,t}}^D =
    {\begin{bmatrix}
        {C_{0,t}}^D \\ {C_{1,t}}^D \\ {C_{2,t}}^D \\ {C_{3,t}}^D
    \end{bmatrix}}^\top
    \begin{bmatrix}
       \beta_{0,3}(\tau)\\ \beta_{1,3}(\tau) \\ \beta_{2,3}(\tau) \\ \beta_{3,3}(\tau)
    \end{bmatrix}
    = \bm{D}_t\bm{\beta}_t
\end{equation}
Where $\bm{D}_t$ and $\bm{\beta}_t$ are vectors of Bernstein coefficients and basis functions for the $t$th interval respectively. Likewise, the forecasted \gls{RES} can also be defined with such vectors. The subscripts of the coefficient $C$ are its indices and its superscript is what the coefficient is describing (here the demand).

\subsubsection{Bernstein Zero-order Continuity} The Bernstein polynomials that describe power during the time interval should be continuous, even at the joint point of the adjutant intervals. To preserve zero-order continuity between intervals, the first Bernstein coefficient of $t$th interval should be equal to the last Bernstein coefficient of the previous interval. Zero-order continuity prevents generation units from jumping in power.
\begin{equation}\label{eq:zero_order}
    {C_{0,t,i}}^p={C_{3,t-1,i}}^p\;\;\;\;\text{\scriptsize $t\in T,\;i\in\mathcal{I}$}
\end{equation}
As \gls{SU} and \gls{SD} trajectories are considered, \cref{eq:zero_order} will be maintained when units start-up or shut down.

\subsubsection{Bernstein First-order Continuity} Furthermore, it is physically impossible to have instant changes in ramping. Accordingly, the differential of the Bernstein polynomials that describe power should also be continuous. According to \cref{eq:ber_def}, this constraint is given by,
\begin{equation}\label{eq:first_order}
    {C_{1,t,i}}^p-{C_{0,t,i}}^p={C_{3,t-1,i}}^p-{C_{2,t-1,i}}^p\;\;\;\;\text{\scriptsize $t\in T,\;i\in\mathcal{I}$}
\end{equation}
Again, as \gls{SU} and \gls{SD} trajectories are considered, \cref{eq:first_order} will be maintained when units start-up or shut down.

\subsubsection{Power Balance}  
The sum of the generation should meet the demand over the time horizon. According to \cref{sec:sum_diff} we can sum over the coefficients of the Bernstein polynomials that represent the generation. The power balance for each hour can be written as,
\begin{equation}\label{eq:power_balance_cont}
    \sum_{i\in\mathcal{I}}(\bm{P}_{t,i}) + {\bm{P}_t}^{\text{\tiny RES}}-{\bm{P}_t}^{\text{curt}}=\bm{D}_t
\end{equation}
Where $\bm{P}$, $\bm{P}^{\text{\tiny RES}}$, and $\bm{P}^{\text{curt}}$ are vectors of Bernstein coefficients for thermal generation, \gls{RES} generation, and \gls{RES} curtailment, respectively. The equivalent of \cref{eq:power_balance_cont} in term of the coefficients will be,
\begin{equation}
    \sum_{i\in\mathcal{I}}({C_{b,t,i}}^p)+{C_{b,t}}^\text{\tiny RES}-{C_{b,t}}^\text{curt}={C_{b,t}}^D\;\;\;\;\text{\scriptsize $t\in T,\;i\in\mathcal{I},\;\forall b$}
\end{equation}

\subsubsection{SU and SD trajectories}
\begin{figure}[!htbp]
    \centering
    \includegraphics[width=0.8\linewidth]{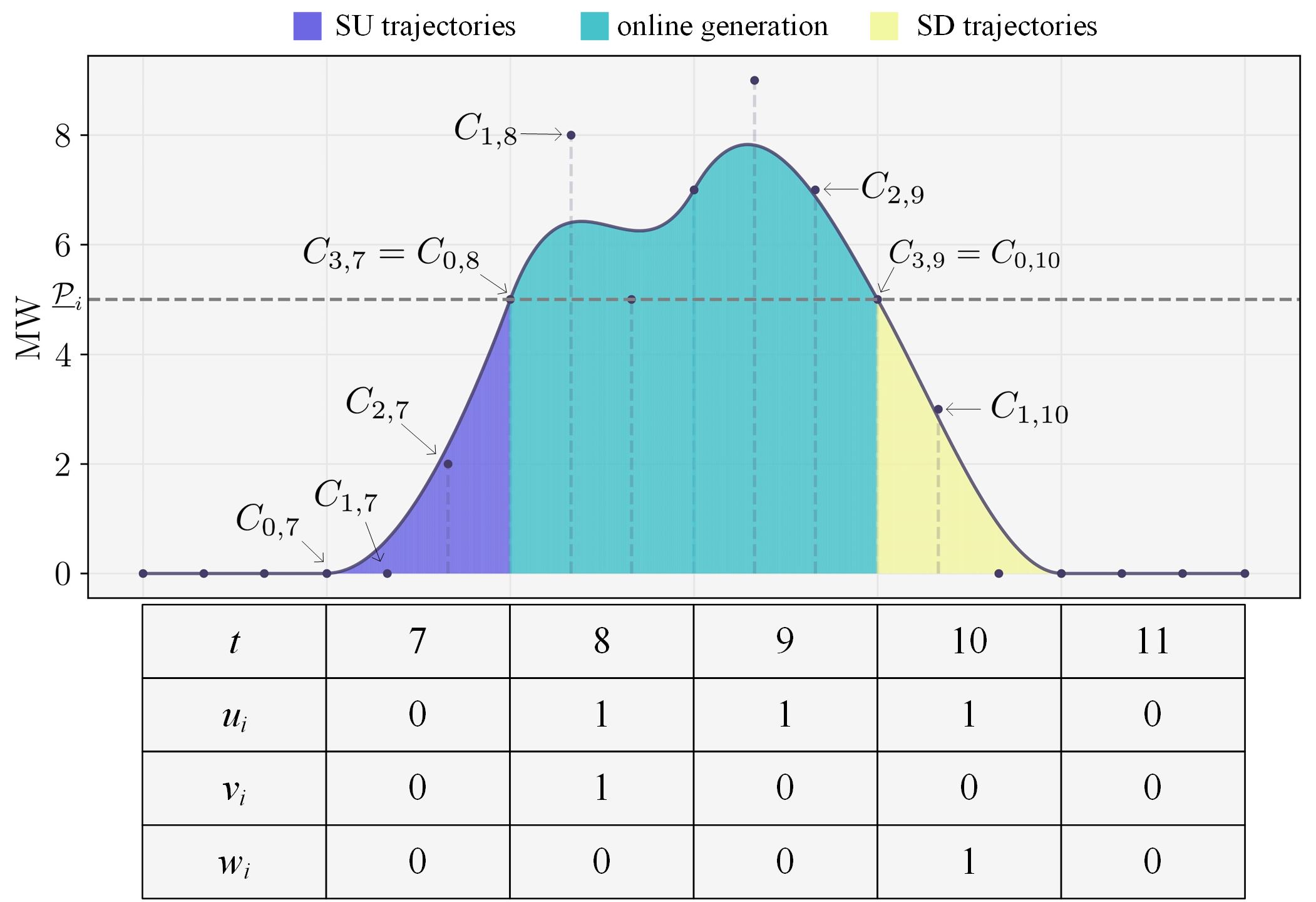}
    \caption{An illustrative sequence of binary variables, SU, and SD trajectories for unit $i$.}
    \label{fig:example}
\end{figure}

As mentioned before, the thermal generation also includes the \gls{SU} and \gls{SD} trajectories. Considering \gls{SU} and \gls{SD} trajectories allows having a fully continuous representation of power generation throughout the time horizon. A sample sequence is illustrated in \cref{fig:example} to show the \gls{SU} and \gls{SD} trajectories for a unit with $\text{SU}^T_i=\text{SD}^T_i=1$h, the binary variable, and the corresponding Bernstein coefficients. In \cref{fig:example}, for instance $C_{0,7}$ is the first Bernstein coefficient when $t=7$. For the sake of brevity, the superscript of coefficient $C$ describing power outputs is not shown.

\subsubsection{Generation Capacity} According to \cref{eq:limit}, the following inequality would be enough to preserve the capacity of units within their limits,
\begin{equation}\label{eq:gen_lim_cont}
    \underline{\mathcal{P}}_i\leq {C_{b,t,i}}^p \leq \overline{\mathcal{P}}_i \;\;\;\;\text{\scriptsize $t\in T,\;i\in\mathcal{I},\;\forall b$}
\end{equation}
Where $b$ is the index of Bernstein coefficients.

\subsubsection{Ramping Capacity}

Using \cref{eq:ber_def}, the following equation is concluded for the differential of a Bernstein polynomial of order three,
\begin{equation}
    {B'_{3,t,i}}=3
    {\begin{bmatrix}
    {C_{1,t,i}}^p-{C_{0,t,i}}^p \\ {C_{2,t,i}}^p-{C_{1,t,i}}^p \\ {C_{3,t,i}}^p-{C_{2,t,i}}^p
    \end{bmatrix}}^\top
    \begin{bmatrix}
        \beta_{0,2}(\tau)\\ \beta_{1,2}(\tau) \\ \beta_{2,2}(\tau)
    \end{bmatrix}
\end{equation}
Which is defined by Bernstein's basis functions of order 2. By referring to \cref{eq:limit} the ramping constraint can be written as,
\begin{equation}\label{eq:ramp_lim_cont}
    \underline{\mathcal{R}}_i\leq {C_{b,t,i}}^p-{C_{b-1,t,i}}^p \leq \overline{\mathcal{R}}_i \;\;\;\;\text{\scriptsize $t\in T,\;i\in\mathcal{I},\;b\in\{1,2,3\}$}
\end{equation}

\subsection{CFCUC Formulation Additions}\label{sec:CFCUC}

\subsubsection{RoCoF constraint}

The \gls{RoCoF} constraint is straightforward, as it's linearly derived from the well-known swing equation. Assuming that the power of the lost unit is described with a vector of Bernstein coefficients, $\bm{P}_{t,\ell}$, according to \cref{eq:limit} the following constraint will be enough to ensure \gls{RoCoF} restrictions,
\begin{equation}\label{eq:rocof}
\bm{P}_{t,\ell} \leq \frac{2\Delta f_{\text{crit}}^{'}}{f_0}\sum_{i\in \mathcal{I}, i\neq \ell}(H_i\mathcal{M}_i u_{t,i})\;\;\text{\scriptsize $t\in T,\;\forall \ell$}
\end{equation}
Where $\ell$ is the index of the lost generator, $\Delta f_{crit}^{'}$ is the critical \gls{RoCoF}, $H$ is the inertia, and $\mathcal{M}$ is the base power of the unit. Note that $\bm{P}$ contains the Bernstein coefficients.

\subsubsection{Steady State Frequency Constraint}\label{sec:qss}

A linear steady-state frequency constraint can be derived from the swing equation. It is assumed that frequency is converged and there has been enough time for units to deliver their reserve power. To make sure that the steady state frequency is not violated, this constraint can be derived from the swing equation,
\begin{equation}\label{eq:ssf}
\sum\limits_{i \in \mathcal{I}, i\neq \ell}\bm{R}_{t,i}\geq \bm{P}_{t,\ell}-D\bm{D}_t\Delta f^{\text{ss}}_{\text{crit}}\;\;\text{\scriptsize $t\in T,\;\forall \ell$}
\end{equation}
This constraint is very similar to the reserve constraint traditionally used in the \gls{UC} problem.

\subsubsection{Frequency Nadir Constraint}

Unlike the previous two constraints, the frequency nadir is more complicated to drive from the swing equation. An essential assumption to drive the frequency nadir is that the units can deliver their headroom linearly and in $T^g$ seconds \cite{badesa2019simultaneous,trovato2018unit}. The continuous representation of the frequency nadir is given by,
\begin{equation}\label{eq:nadir}
    f^{\text{nadir}}(\tau)=\ddfrac{f_0T^g p_\ell(\tau)^2}{4r_\ell(\tau)H_\ell - DT^gf_0\mathcal{D}(\tau) p_\ell(\tau)}
\end{equation}
where $H_\ell$ is the available weighted inertia after outage. Note that $H_\ell$ is only a variable of $u$ and is constant during each hourly interval. Additional information regarding the derivation of \cref{eq:nadir} from the swing equation is elaborated upon in \cite{ferrandon2022inclusion}.

\begin{figure}[!htbp]
    \centering
    \includegraphics[width=1\linewidth]{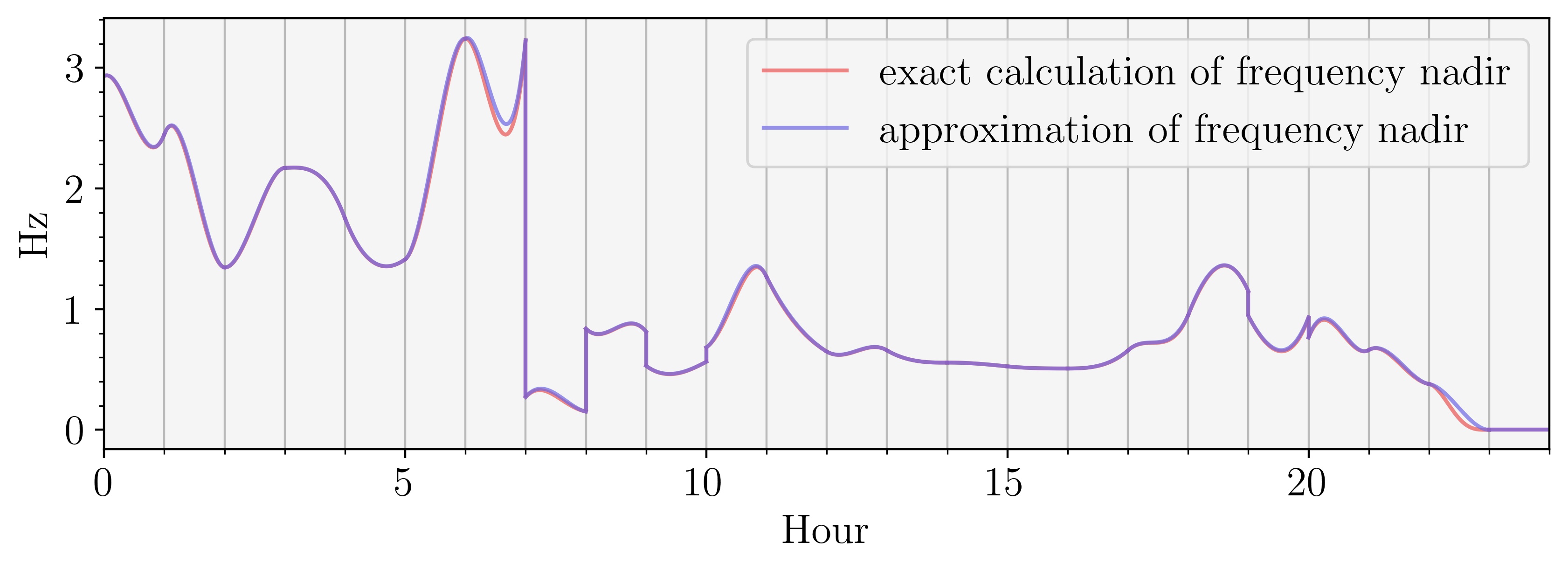}
    \caption{The exact calculation of \cref{eq:nadir} is compared with the approximation in \cref{eq:nadir_approx}}
    \label{fig:accuracy}
\end{figure}

Assuming that all of the continuous elements in \cref{eq:nadir} are described by Bernstein polynomials of order 3 in hourly intervals, the term in the numerator includes a square of a Bernstein polynomial of order 3, which leads to a Bernstein polynomial of order 6 (see \cref{sec:multi_poli}). The first term in the denominator is a product of a Bernstein polynomial of order 3 in a constant for each interval (remember that we assumed $u$ can only change at the beginning of the hourly intervals). The second term in the denominator includes a product of the Bernstein polynomial of order 3 into another Bernstein polynomial of order 3, which again leads to a Bernstein polynomial of order 6. To calculate the denominator, the order of the first term should be elevated based on \cref{sec:degree_elev} from 3 to 6. Then the frequency nadir will be described by a fraction, where nominator and denominator are Bernstein polynomials of order 6. The coefficients can be calculated based on the coefficients of the initial Bernstein polynomials of order 3, but that will cause a temporal dependency i.e. some of the coefficients will be dependent on the previous or the following control points. This is very hard to calculate, let alone the non-linearities, and the process of applying it to the \gls{MILP} problem at hand. Let $p_\ell(\tau)={B_3}^p$, $r_\ell(\tau)={B_3}^r$, and $\mathcal{D}(\tau)={B_3}^D$ (we defined continuous variables in \cref{eq:nadir} as Bernstein polynomials of order 3). To simplify the calculation of frequency nadir we make these assumptions,
\begin{subequations}
    \begin{align}\label{eq:approx1}
    &({B_3}^p)^2\approx
    \begin{bmatrix}
        ({C_0}^p)^2 & ({C_1}^p)^2 & ({C_2}^p)^2 & ({C_3}^p)^2
    \end{bmatrix}
    \bm{\beta}_t
    \\\label{eq:approx2}
    &{B_3}^D{B_3}^p\approx
    \begin{bmatrix}
        {C_0}^D{C_0}^p & {C_1}^D{C_1}^p & {C_2}^D{C_2}^p & {C_3}^D{C_3}^p
    \end{bmatrix}
    \bm{\beta}_t
    \end{align}
\end{subequations}
These assumptions eliminate the temporal dependency and let us define the nominator and denominator of \cref{eq:nadir} as Bernstein polynomials of order 3, like any other continuous variable in this paper. Note that for the sake of brevity, the indices of time and units are not shown in \cref{eq:approx1,eq:approx2}. The coefficients of the approximated frequency nadir will be calculated as,
\begin{equation}\label{eq:nadir_approx}
    {C_{b,t,\ell}}^f=\ddfrac{f_0T^g ({C_{b,t,\ell}}^p)^2}{4{C_{b,t,\ell}}^r H_{t,\ell} - DT^gf_0{C_{b,t,\ell}}^D {C_{b,t,\ell}}^p}
\end{equation}
Both exact calculation and approximation of frequency nadir are shown in \cref{fig:accuracy}. \Cref{fig:accuracy} shows the frequency nadir caused by the outage of a sample unit throughout the day and confirms that the approximation in \cref{eq:approx1,eq:approx2} is reasonable.

Considering that even the approximated version of \cref{eq:nadir} is non-linear and applying it to the problem at hand would be challenging, we propose a data-driven method to drive the frequency nadir constraint. The dataset is synthetically built, using the algorithm proposed in \cite{rajabdorri2023inclusion}. For the purpose of this paper, lost power, online inertia, and available reserve are chosen as features, as they are the variables appearing in \cref{eq:nadir}. To label each sample, the frequency nadir is calculated using \cref{eq:nadir_approx}. Utilizing this labeled dataset, and with the help of a classic machine learning method (regression in this paper), a linear constraint on frequency nadir can be attained. According to \cref{eq:limit}, this constraint will prevent outages in which the frequency nadir will exceed the specified critical value.
\begin{equation}\label{eq:learned_nadir}
    \alpha_0+\alpha_1{C_{b,t,\ell}}^p+ \alpha_2 H_\ell+\alpha_3{C_{b,t,\ell}}^r\leq \Delta f_{\text{crit}}^{\text{nadir}}
\end{equation}
The effectiveness, accuracy, and applicability of a data-driven constraint depend on many factors, including the quality of the dataset in reflecting the behavior of the system, the complexity of learning labels, and the cross-validation score of the learned constraint, among others. Later in the result section, the precision of \cref{eq:learned_nadir} in learning frequency nadir is proven for the case study.

\subsubsection{Summary of the proposed formulation} \label{sec:summary} The proposed continuous \gls{FCUC} is solved with the objective of minimizing \cref{eq:obj}. The binary logic of the problem is similar to \cref{eq:bi1,eq:bi2,eq:up_time,eq:down_time} with minimal changes. The power balance equality is defined in \cref{eq:power_balance}. Zero-order and first-order continuity are necessary to ensure that the scheduled power and ramp are continuous and are assured by \cref{eq:zero_order,eq:first_order}, respectively. Generation and ramping capacity limitations are forced by \cref{eq:gen_lim_cont,eq:ramp_lim_cont}, respectively. The frequency-related constraints considered in this paper are \gls{RoCoF}, steady-state frequency, and frequency nadir constraints. \gls{RoCoF} and steady-state frequency constraints are linearly derived from the swing equation and are presented in \cref{eq:rocof,eq:ssf}. Then it's argued that the continuous frequency nadir is hard to implement in an \gls{MILP} formulation analytically, so a data-driven method is introduced alternatively, leading to \cref{eq:learned_nadir}.

\section{Numerical Results}\label{sec:results}

The proposed methodology is applied to real data of a sample day on the Spanish island of La Palma. La Palma power system comprises 11 generation units and has a peak demand of around 40 MW. As it's a system dominated by fast diesel units, $T^g$ in \cref{eq:nadir} is assumed to be 3 s, and a load damping factor ($D$) of 1\% is assumed. The optimization problem has been written in GAMS and solved by the Gurobi solver \cite{gurobi}. For the sake of comparison, three cases are defined, and the results are compared.
\begin{itemize}
    \item \textbf{\gls{CUC}}: the standard continuous \gls{UC} without the frequency-related constraints similar to the one presented in \cite{parvania2015unit} but with the additional \gls{SU} and \gls{SD} constraints.
    \item \textbf{\gls{RoCoF}-\gls{CUC}}: the proposed \gls{CFCUC} minus the frequency nadir constraint.
    \item \textbf{\gls{CFCUC}}: the complete \gls{CFCUC} formulation summarized in \cref{sec:summary}. Different frequency nadir thresholds are showcased to better demonstrate the effectiveness of the proposed method.
\end{itemize}
All of the inputs, codes, and results of this study are available online. The interested reader is referred to \href{https://doi.org/10.5281/zenodo.10401279}{https://doi.org/10.5281/zenodo.10401279}.

A summary of the obtained results for the three cases is presented in \cref{tab:results}.
\begin{table*}[!htbp]
    \footnotesize \centering
    \caption{Summary of the results}
    \begin{tabular}{c|c|c|c|c}
    case &  $\Delta f^\text{nadir}>2.5$ Hz & operation cost & run-time & score \\
    \toprule
    \gls{CUC}  & $1151'$ & 118.03 k\euro & $82''$& - \\
    \gls{RoCoF}-\gls{CUC}  & $677'$ & 118.54 k\euro & $52''$ & - \\
    \gls{CFCUC} $\Delta f_{\text{crit}}^{\text{nadir}}=3$ Hz & $556'$ & 118.78 k\euro & $77''$ & 99.93\%\\
    \gls{CFCUC} $\Delta f_{\text{crit}}^{\text{nadir}}=2.5$ Hz & $0'$ & 122.48 k\euro & $196''$ & 99.67\%\\
    \gls{CFCUC} $\Delta f_{\text{crit}}^{\text{nadir}}=2$ Hz & $0'$ & 128.5 k\euro& $1725''$ & 99.61\%\\
    \bottomrule
    \end{tabular}
    \label{tab:results}
\end{table*}
The operation cost of \gls{CUC} is less than the other cases, as there are no frequency-related constraints to prevent poor frequency responses. And as expected, the operation cost gets more expensive as the constraints get more restrictive. This is the cost that the operator pays to improve the frequency response.

\begin{figure}

\begin{subfigure}{\linewidth}
\centering
    \includegraphics[clip,width=0.8\linewidth]{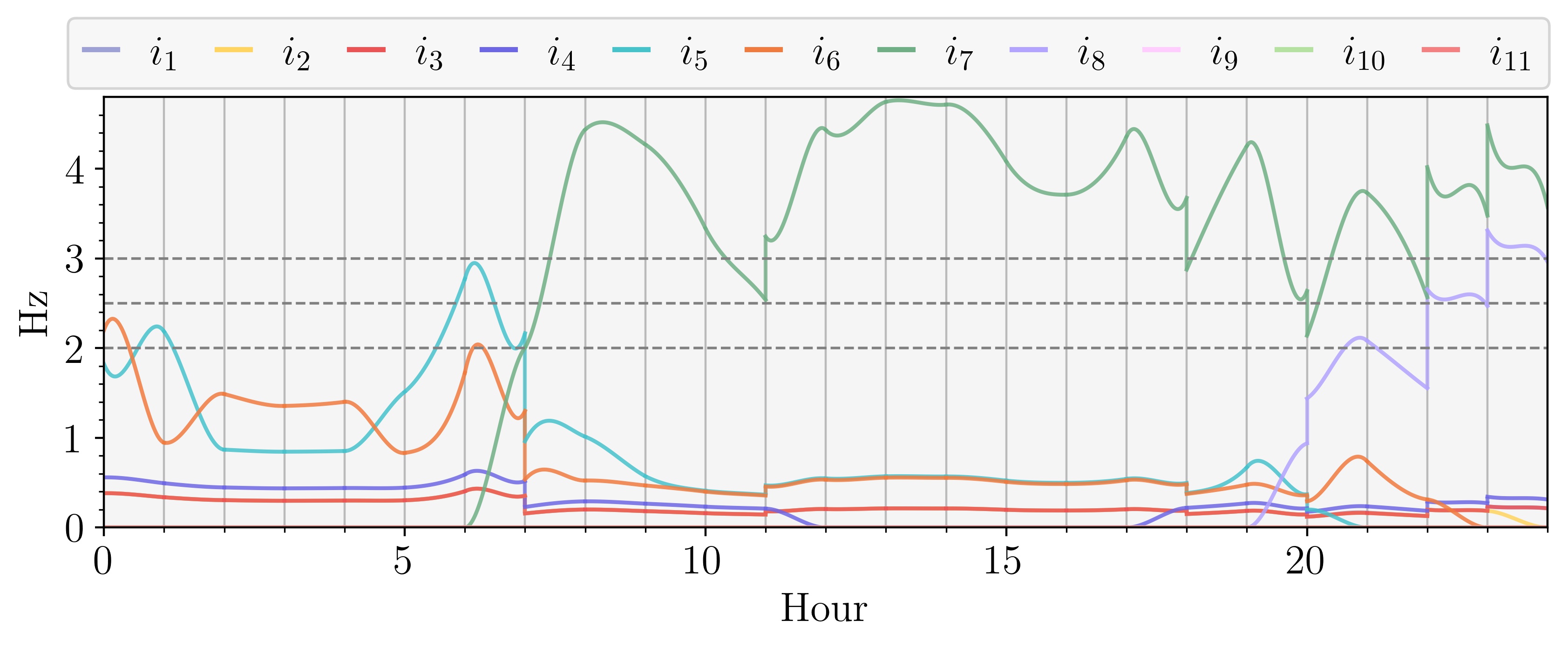}
    \caption{\scriptsize \gls{CUC}}
    \label{fig:fnadir_CUC}
\end{subfigure}
         
\begin{subfigure}{\linewidth}
\centering
    \includegraphics[clip,width=0.8\linewidth]{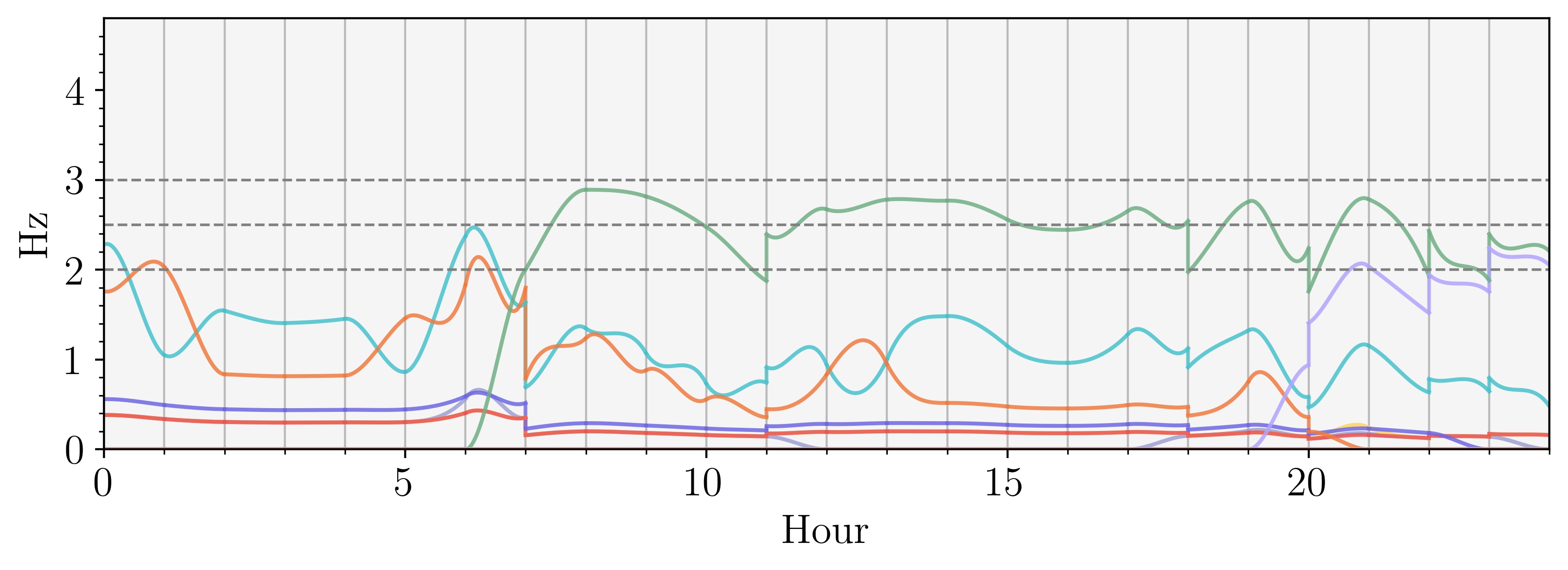}
    \caption{\scriptsize \gls{CFCUC} $\Delta f_{\text{crit}}^{\text{nadir}}=3$ Hz}
    \label{fig:fnadir_5}
\end{subfigure}

\begin{subfigure}{\linewidth}
\centering
    \includegraphics[clip,width=0.8\linewidth]{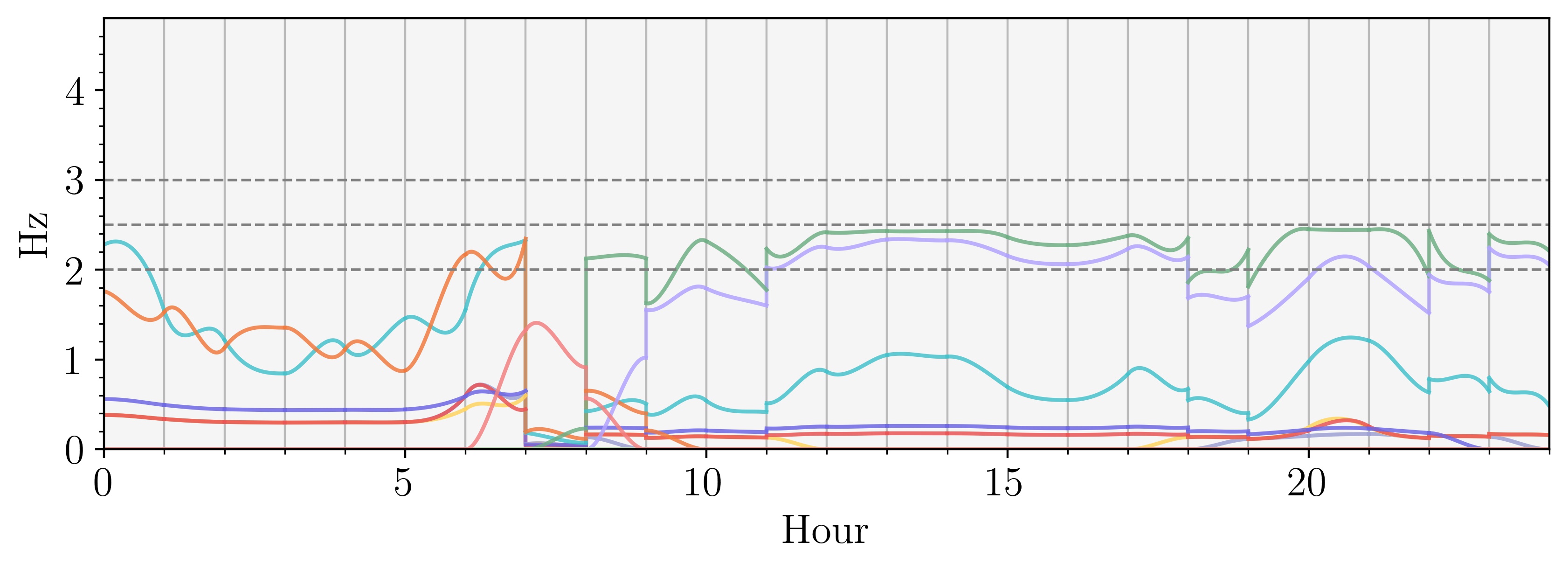}
    \caption{\scriptsize \gls{CFCUC} $\Delta f_{\text{crit}}^{\text{nadir}}=2.5$ Hz}
    \label{fig:fnadir_4}
\end{subfigure}

\begin{subfigure}{\linewidth}
\centering
    \includegraphics[clip,width=0.8\linewidth]{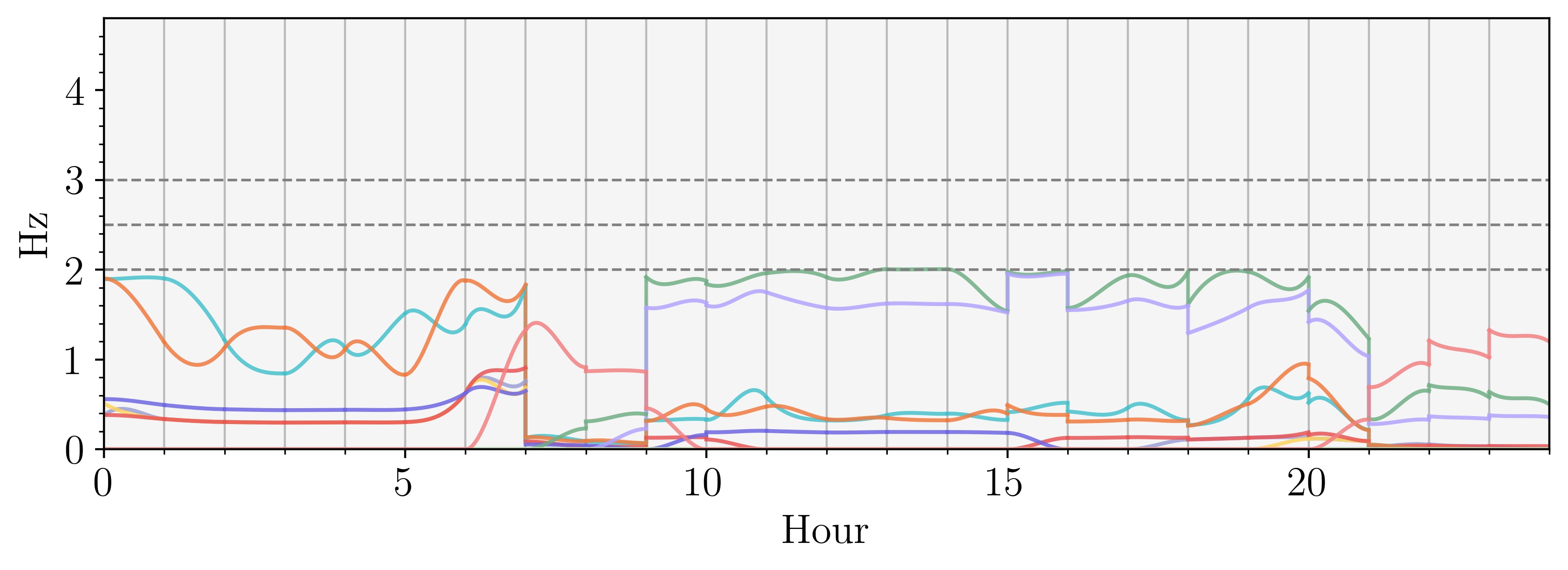}
    \caption{\scriptsize \gls{CFCUC} $\Delta f_{\text{crit}}^{\text{nadir}}=2$ Hz}
    \label{fig:fnadir_3o5}
\end{subfigure}

\caption{Frequency nadir calculated by \cref{eq:nadir} of the units $i_1$ to $i_{11}$ committed by (a) \gls{CUC}, (b) \gls{CFCUC} with $\Delta f_{\text{crit}}^{\text{nadir}}=3$ Hz, (c) \gls{CFCUC} with $\Delta f_{\text{crit}}^{\text{nadir}}=2.5$ Hz, (d) \gls{CFCUC} with $\Delta f_{\text{crit}}^{\text{nadir}}=2$ Hz.}
\label{fig:fnadir}
\end{figure}

Despite the complexities of the continuous formulation and the frequency constraints, especially the frequency nadir, the proposed \gls{CFCUC} is solved in a timely manner. This is mainly because of considering the \gls{SU} and \gls{SD} trajectories, which simplifies many parts of the continuous formulation, and also estimating the frequency nadir threshold with the linear constraint in \cref{eq:learned_nadir}.
The cross-validated accuracy of the learning process (logistic regression) is depicted in \cref{tab:results} within the score column. In this process, 70\% of the data samples serve as the training dataset, while the remaining 30\% comprise the test dataset. The score reflects the accuracy of applying the learned estimation from the training dataset to the test dataset. Notably, the score exhibits a very high value, indicating that \cref{eq:learned_nadir} can effectively identify operation points leading to a frequency nadir deviation exceeding $\Delta f_{\text{crit}}^{\text{nadir}}$ with exceptional precision.
Later, it's assured in \cref{fig:fnadir} that the proposed \gls{CFCUC} has been able to flawlessly prevent frequency nadir deviating from $\Delta f_{\text{crit}}^{\text{nadir}}$.

In the second column, the number of minutes that the outage of a unit would lead to $\Delta f^\text{nadir}>2.5$ Hz is calculated. In the \gls{CUC} generation schedule during $1151$ minutes through the day, there is a unit online whose outage will cause a frequency deviation higher than $2.5$ Hz. The number of minutes reduces as the constraints get more restrictive. And when $\Delta f_{\text{crit}}^{\text{nadir}}>2.5$ Hz is imposed, the frequency deviation is never higher than $2.5$ Hz.

\begin{figure}[!htbp]
\begin{subfigure}{\linewidth}
    \includegraphics[clip,width=\linewidth]{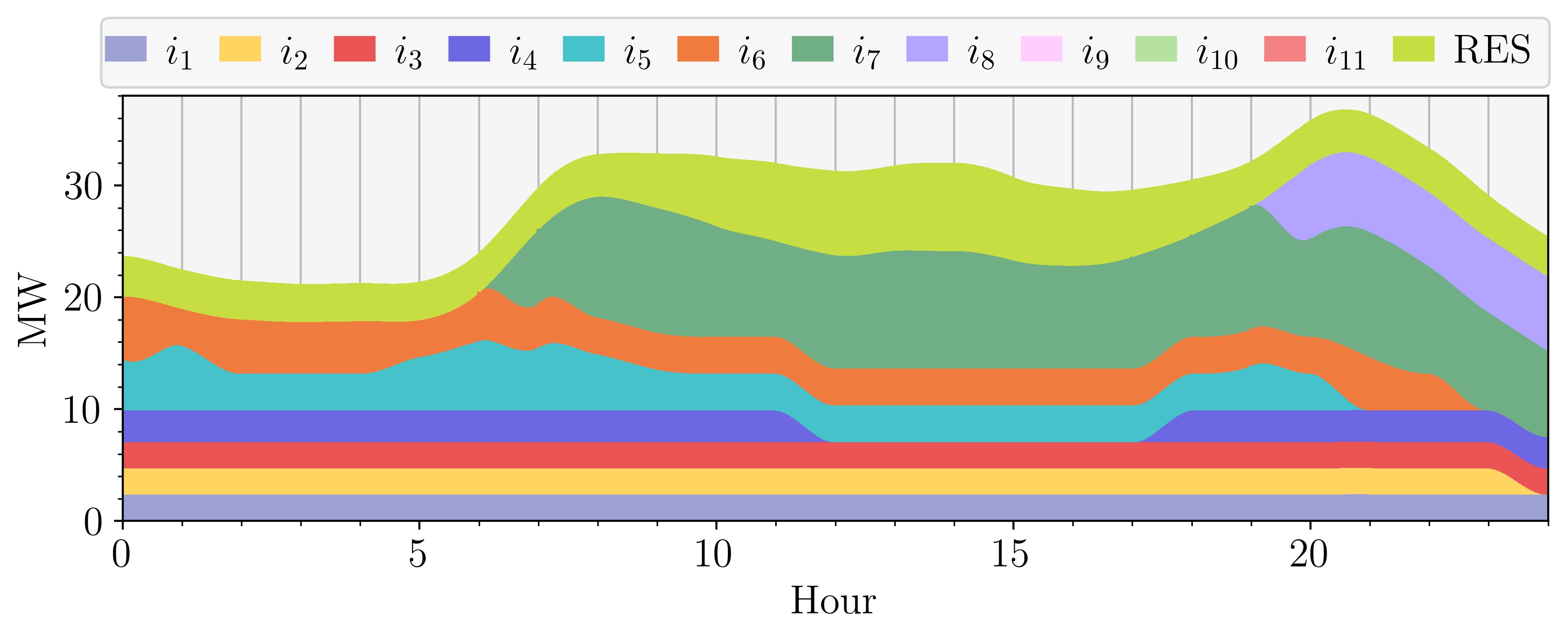}
    \caption{\scriptsize \gls{CUC}}
    \label{fig:gen_CUC}
\end{subfigure}

\begin{subfigure}{\linewidth}
    \includegraphics[clip,width=\linewidth]{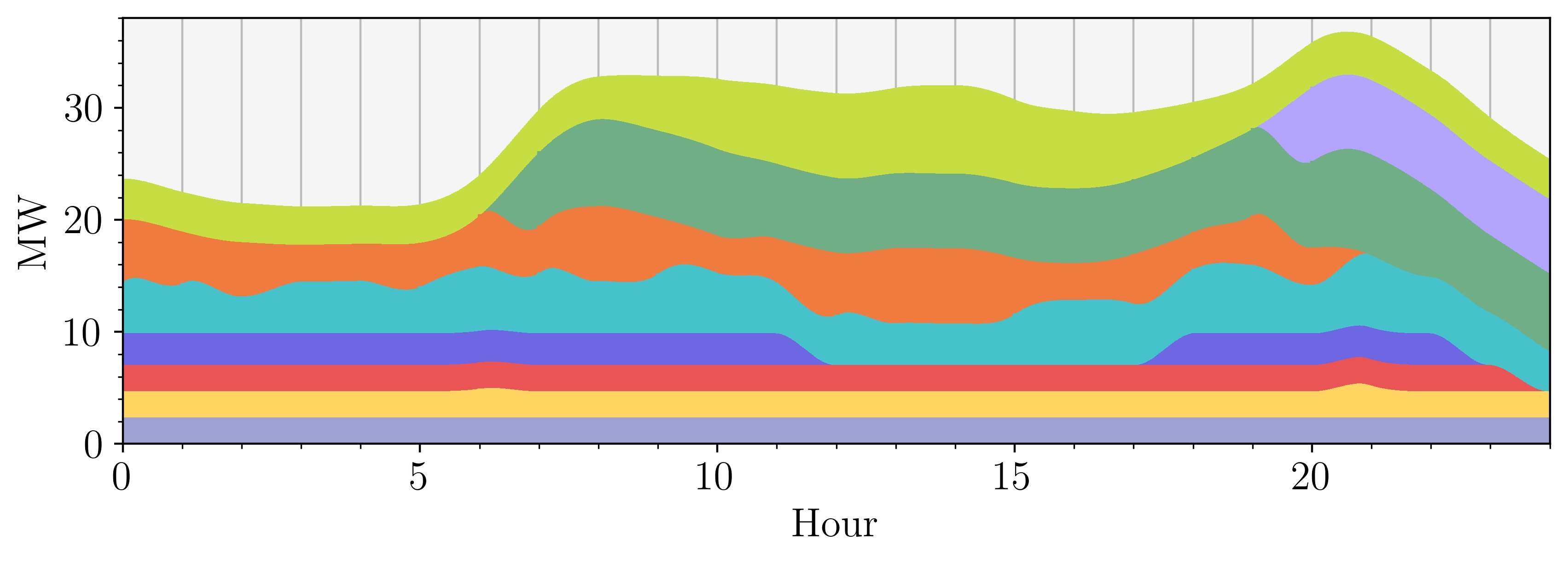}
    \caption{\scriptsize \gls{RoCoF}-CUC}
    \label{fig:gen_RoCoF_UC}
\end{subfigure}

\begin{subfigure}{\linewidth}
    \includegraphics[clip,width=\linewidth]{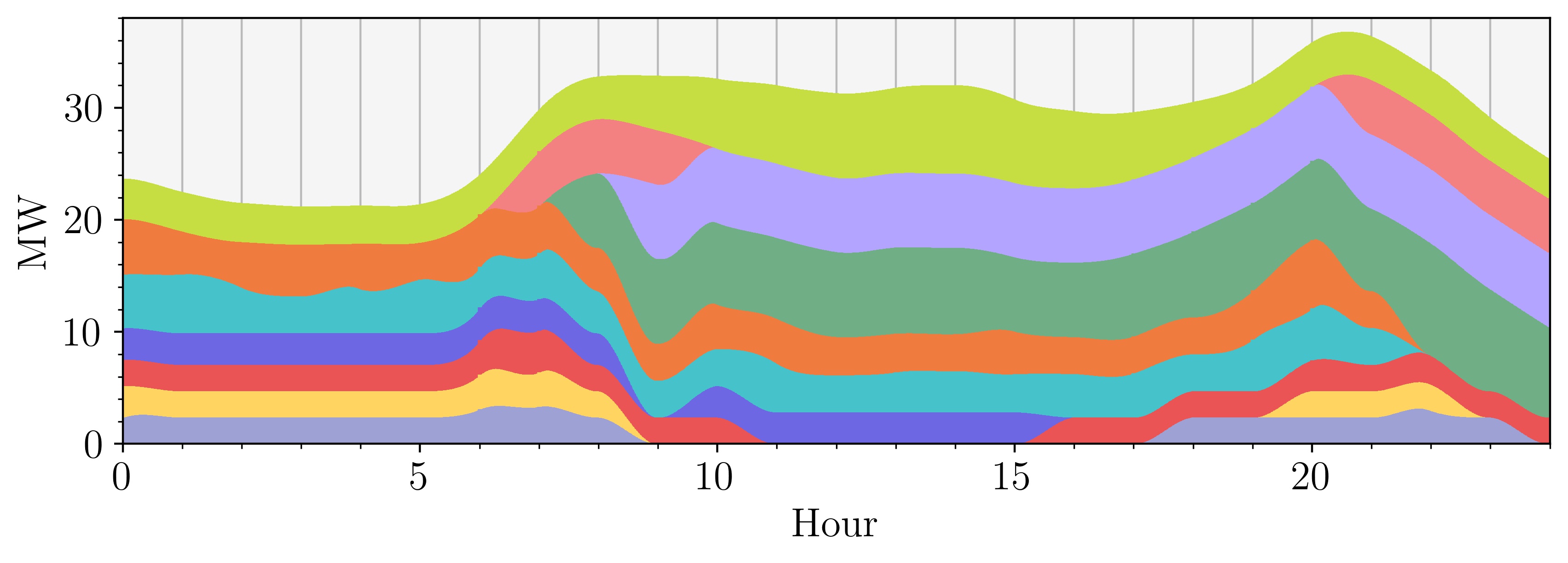}
    \caption{\scriptsize \gls{CFCUC} $\Delta f_{\text{crit}}^{\text{nadir}}=2$ Hz.}
    \label{fig:gen_CFCUC}
\end{subfigure}

\caption{Generation schedule of \gls{RES} and the units $i_1$ to $i_{11}$ for (a) \gls{CUC}, (b) \gls{RoCoF}-CUC, and (c) \gls{CFCUC} when $\Delta f_{\text{crit}}^{\text{nadir}}$ is $2$ Hz }
\end{figure}

The scheduled generation for \gls{CUC}, \gls{RoCoF}-\gls{CUC}, and \gls{CFCUC} are shown in \cref{fig:gen_CUC,fig:gen_RoCoF_UC,fig:gen_CFCUC}.
The generation schedule for \gls{CUC} and \gls{RoCoF}-\gls{CUC} are very similar, but it seems when the \gls{RoCoF} constraint is added, the generation of units 5 and 6 is increased to reduce the generation of unit 7, to prevent the big outages that were violating \gls{RoCoF} limitation.
Then for the \gls{CFCUC} with a restricting $\Delta f_{\text{crit}}^{\text{nadir}}$ of $2$ Hz, the combination of online units further changes to keep the frequency nadir deviation below the critical value throughout the day, by committing units with higher inertia and by preventing relative big power outages.

To better compare how the proposed \gls{CFCUC} has been able to improve the frequency response by preventing poor frequency incidents from happening, the exact calculated frequency nadir from \cref{eq:nadir} for all the online units throughout the day is shown in \cref{fig:fnadir}.
It can be seen in \cref{fig:fnadir,fig:fnadir_5,fig:fnadir_4,fig:fnadir_3o5} how the frequency nadir constraint in \cref{eq:learned_nadir} has been able to perfectly prevent frequency nadir deviation to go over $\Delta f_{\text{crit}}^{\text{nadir}}$, although the frequency nadir calculated in \cref{eq:nadir} has been approximated once with \cref{eq:approx1} and \cref{eq:approx2} and then estimated with a learning process in \cref{eq:learned_nadir}. This further proves the accuracy of approximation of \cref{eq:approx1} and \cref{eq:approx2} (as showed also before in \cref{fig:accuracy}) and the machine learning method that is used. Something else that is evident in \cref{fig:fnadir} is that intra-hour changes in the frequency nadir deviation can be substantial. In some hours the deviation can differ by more than 2 Hz during the same hour. Something that has been ignored in the traditional discrete \gls{FCUC}. For example in \gls{CUC} between hours 7 and 8, the schedule generation of unit $i_7$ increases a lot, so the frequency nadir deviation at the end of the hour is around 2.5 Hz more comparing to the beginning of the hour if the unit fails.

\section{Conclusion}\label{sec:conc}
This paper introduces a \gls{CFCUC} approach designed to mitigate the impact of intra-hour real-time power fluctuations on system frequency, thereby preventing undesirable frequency responses that could fall below critical thresholds. The presented results demonstrate the efficacy of the proposed model, showcasing its ability to be solved efficiently while achieving the intended goal of maintaining the quality of frequency response. This approach represents a significant advancement in addressing the unit commitment problem within a continuous-time framework, offering improved accuracy and computational efficiency in handling frequency-related constraints. The utilization of a data-driven frequency nadir constraint method in this study proves effective, affirming the reasonability of the simplifying assumptions adopted. As the proposed data-driven frequency nadir constraint is effective, while not imposing much computational burden on the \gls{UC} problem, its usage is recommended, as a substitute for analytical methods that require extensive linearization.

This research contributes valuable insights into ensuring reliable and stable power system operations against dynamic real-time changes. The next step for this line of research can be applying the method to bigger inertia systems, and also considering under-frequency load-shedding, which is much more challenging to implement in an \gls{MILP} formulation.



 \bibliographystyle{elsarticle-num} 
 \bibliography{cas-refs}





\end{document}